\documentclass[11pt]{revtex4}
\usepackage{amssymb,epsf}
\usepackage{latexsym}

\begin{document}

\title{Thermodynamic properties of asymptotically Reissner-Nordstr\"{o}m
black holes}
\author{S. H. Hendi\footnote{email address: hendi@shirazu.ac.ir}}
\affiliation{Physics Department and Biruni Observatory, College of Sciences, Shiraz
University, Shiraz 71454, Iran \\
Research Institute for Astrophysics and Astronomy of Maragha (RIAAM), P.O.
Box 55134-441, Maragha, Iran}

\begin{abstract}
Motivated by possible relation between Born-Infeld type nonlinear
electrodynamics and an effective low-energy action of open string
theory, asymptotically Reissner--Nordstr\"{o}m black holes whose
electric field is described by a nonlinear electrodynamics (NLED)
are studied. We take into account a four dimensional topological
static black hole ansatz and solve the field equations, exactly,
in terms of the NLED as a matter field. The main goal of this
paper is investigation of thermodynamic properties of the obtained
black holes. Moreover, we calculate the heat capacity and find
that the nonlinearity affects the minimum size of stable black
holes. We also use Legendre-invariant metric proposed by Quevedo
to obtain scalar curvature divergences. We find that the
singularities of the Ricci scalar in Geometrothermodynamics (GTD)
method take place at the Davies points.
\end{abstract}

\maketitle


\section{Introduction}

Recent developments on the nonlinear generalization of Maxwell
electrodynamics showed that Born-Infeld (BI) type NLED may be arisen as a
low energy limit of heterotic string theory. These recent progresses on the
NLED theories indicated that we should consider quartic corrections of
Maxwell field strength in the action of abelian electrodynamics \cite%
{Kats1,Kats2,Kats3,Kats4,Kats5,Kats6,Kats7}, which led to an increased
interest for BI type NLED theories.

Although NLED has been applied to various theories of black holes,
one can find its interesting applications of very compact
astrophysical objects. Indeed, the effects of NLED theories become
quite important in super-strongly magnetized compact objects, such
as pulsars, particular neutron stars, magnetars and strange quark
magnetars \cite{Cuesta1,Cuesta2,Bialynicka}. Also, it has been
shown that, NLED modifies in a fundamental basis the concept of
gravitational redshift and its dependency of any background
magnetic field as compared to the well-established method
introduced by standard general relativity. Furthermore, it has
been recently shown that NLED theories can remove both of the big
bang and black hole singularities \cite{Cuesta3,Salim1,Cuesta4}.

Besides NLED applications, we should note that BI type theories of
are special among other classes of NLED by their remarkable
properties such as birefringence phenomena, free of the shock
waves \cite{Boillat1,Boillat2} and also enjoying an
electric-magnetic duality \cite{Boillat3}. Furthermore,
considering the relation between AdS/CFT correspondence and
superconductivity phenomenon, it was shown that the BI type
theories make an crucial effect on the condensation, the critical
temperature and energy gap of the superconductors \cite{JJing}.

In this paper, we investigate thermodynamic properties of the
recently proposed black hole solutions of BI type models
\cite{HendiAnn}. Moreover, to better understand the role of NLED,
we investigate thermodynamic stability of the black hole solutions
in canonical ensemble. It helps us to have a deep perspective to
study how the nonlinearity affects the thermodynamic behavior of
the black holes.

The plan of the paper is as follows. In Sec. \ref{Sol}, we give a brief note
of the four dimensional topological static black hole solutions. Section \ref%
{Therm} is devoted to calculation of conserved and thermodynamics quantities
and investigation of the first law of thermodynamics. Then, we study the
thermodynamics stability in canonical ensemble and show that the
nonlinearity affects the value of the horizon radius for stable black holes.
This paper ends with some conclusions.

\section{Black hole solutions with Reissner-Nordstr\"{o}m asymptote}

\label{Sol}

In Ref. \cite{HendiAnn}, $4$-dimensional solutions of two classes of NLED
coupled to gravity was considered. This model is described by the following
field equations
\begin{equation}
G_{\mu \nu }+\Lambda g_{\mu \nu }=\alpha \left( \frac{1}{2}g_{\mu \nu }L(%
\mathcal{F})-2F_{\mu \lambda }F_{\nu }^{\;\lambda }L_{\mathcal{F}}\right) ,
\label{FE1}
\end{equation}
\begin{equation}
\partial _{\mu }\left( \sqrt{-g}L_{\mathcal{F}}F^{\mu \nu }\right) =0,
\label{FE2}
\end{equation}
where $G_{\mu \nu }$ and $\Lambda $ are, respectively, the
Einstein tensor and the negative cosmological constant,
$L_{\mathcal{F}}=\frac{dL(\mathcal{F})}{d\mathcal{F}}$. As a
source of Einstein gravity, $L(\mathcal{F})$, we consider the
\textit{Exponential form of Nonlinear Electromagnetic Field}
(ENEF) and the \textit{Logarithmic form of Nonlinear
Electromagnetic Field} (LNEF), in which their Lagrangians are
\begin{equation}
L(\mathcal{F})=\left\{
\begin{array}{ll}
\beta ^{2}\left( \exp (-\frac{\mathcal{F}}{\beta ^{2}})-1\right) ,
& \text{ENEF} \\
-8\beta ^{2}\ln \left( 1+\frac{\mathcal{F}}{8\beta ^{2}}\right) ,
& \text{LNEF}
\end{array}
\right. .  \label{LNon}
\end{equation}
The spherically symmetric solutions with a radial electric field
were found in \cite{HendiAnn}. Here, we extend the solutions of
Ref. \cite{HendiAnn} to the general solutions with various horizon
topologies. General static solutions may be described by the
following line element
\begin{eqnarray}
ds^{2}
&=&-N(r)f(r)dt^{2}+\frac{dr^{2}}{f(r)}+r^{2}\breve{g}_{ij}dx^{i}dx^{j},
\label{Metric} \\
\breve{g}_{ij}dx^{i}dx^{j} &=&\left\{
\begin{array}{cc}
d\theta ^{2}+\sin ^{2}\theta d\phi ^{2} & k=1 \\
d\theta ^{2}+\sinh ^{2}\theta d\phi ^{2} & k=-1 \\
d\theta ^{2}+d\phi ^{2} & k=0%
\end{array}%
\right. ,  \label{gij}
\end{eqnarray}%
with
\begin{equation}
f(r)=k-\frac{2M}{r}-\frac{\Lambda r^{2}}{3}+\left\{
\begin{array}{ll}
\frac{\beta Q}{3\sqrt{L_{W}}}\left[ 1+L_{W}+\frac{4}{5}L_{W}^{2}\digamma {%
\left( \left[ 1\right] ,\left[ \frac{9}{4}\right] ,\frac{L_{W}}{4}\right) }%
\right] , & \text{ENEF}\vspace{0.2cm} \\
\frac{16Q^{2}\digamma {\left( \left[ \frac{1}{2},\frac{1}{4}\right] ,\left[
\frac{5}{4}\right] ,1-\Gamma ^{2}\right) }}{9r^{2}}+\frac{4\beta ^{2}r^{2}}{9%
}\left[ 3\ln \left( \frac{1+\Gamma }{2}\right) +5\left( 1-\Gamma \right) %
\right] , & \text{LNEF}%
\end{array}%
\right. ,  \label{F(r)}
\end{equation}
\begin{equation}
N(r)=C,
\end{equation}
and
\begin{equation}
F_{tr}=-F_{rt}=\frac{Q}{r^{2}}\times \left\{
\begin{array}{ll}
e^{-\frac{L_{W}}{2}}, & \text{ENEF} \\
\frac{2}{\Gamma +1}, & \text{LNEF}%
\end{array}%
\right. ,  \label{Ftr}
\end{equation}%
where $\breve{g}_{ij}dx^{i}dx^{j}$ denotes the metric of two dimensional
hypersurface at $r=$constant and $t=$constant with constant curvature $2k$
in which $k$ is the horizon curvature constant, the integration constants $Q$
and $M$ are the total charge and mass of spacetime, respectively and
\begin{eqnarray*}
\Gamma &=&\sqrt{1+\frac{Q^{2}}{r^{4}\beta ^{2}}}, \\
L_{W} &=&LambertW\left( \frac{4Q^{2}}{\beta ^{2}r^{4}}\right) .
\end{eqnarray*}
Calculations of obtaining the metric functions and electromagnetic
tensor have expressed in Ref. \cite{HendiAnn}. Here, we choose
$N(r)=C=1$ without loss of generality. These solutions
characterize black hole configurations with various horizon
properties depending on the values of the nonlinearity parameter
$\beta $
\cite{HendiJHEP,HendiAnn}. Using the series expansion for large distance ($%
r>>1$), it has been shown that these solutions are asymptotically
Reissner-Nordstr\"{o}m \cite{HendiAnn}. Since the geometric properties of
the solutions were discussed in Ref. \cite{HendiAnn}, in this paper we focus
on the thermodynamics properties as well as stability conditions.

\section{Thermodynamics properties and stability \label{Therm}}

We compute the entropy, temperature and electric potential of the
solutions in order to determine the satisfaction of the first law
of thermodynamics. The entropy of Einsteinian black holes obeys
the Bekenstein-Hawking entropy area law
\cite{Bekenstein1,Bekenstein2,Bekenstein3,Hawking1,Hawking2,Hawking3,Hawking4},
so the entropy associated with the event horizon is one quarter of
its area, i.e.,
\begin{equation}
S=\pi r_{+}^{2}.  \label{Entropy}
\end{equation}
Although the (nonlinear) electromagnetic source changes the values
of inner and outer horizons of charged black objects, it does not
alter the areal law (see Ref. \cite{Gauntlett} for more details).
Since obtained solutions are asymptotically adS, one can prove
this claim by the use of the Gibbs--–Duhem relation to calculate
the entropy with the following approach
\begin{equation}
S=\frac{1}{T}(M-QU)-I, \label{GD}
\end{equation}
where $I$ is the finite total action evaluated by the use of
counterterm method on the classical solutions. After cumbersome
calculations, one can confirm the area law for our solutions.

The electric potential $U$, measured at infinity with respect to
the event horizon, is \cite{Gub1,Gub2}
\begin{equation}
U=\left\{
\begin{array}{ll}
\frac{\beta r_{+}\sqrt{L_{W_{+}}}}{2}\left[ 1+\frac{L_{W_{+}}}{5}\digamma {%
\left( \left[ 1\right] ,\left[ \frac{9}{4}\right] ,\frac{L_{W_{+}}}{4}%
\right) }\right] , & \text{ENEF}\vspace{0.1cm} \\
\frac{2Q}{3r_{+}}\left[ 2\digamma {\left( \left[ \frac{1}{4},\frac{1}{2}%
\right] ,\left[ \frac{5}{4}\right] ,1-\Gamma _{+}^{2}\right) -}\frac{1}{%
\left( 1+\Gamma _{+}\right) }\right] , & \text{LNEF}%
\end{array}%
\right. .  \label{U}
\end{equation}

\begin{figure}[tbp]
$%
\begin{array}{cc}
\epsfxsize=8cm \epsffile{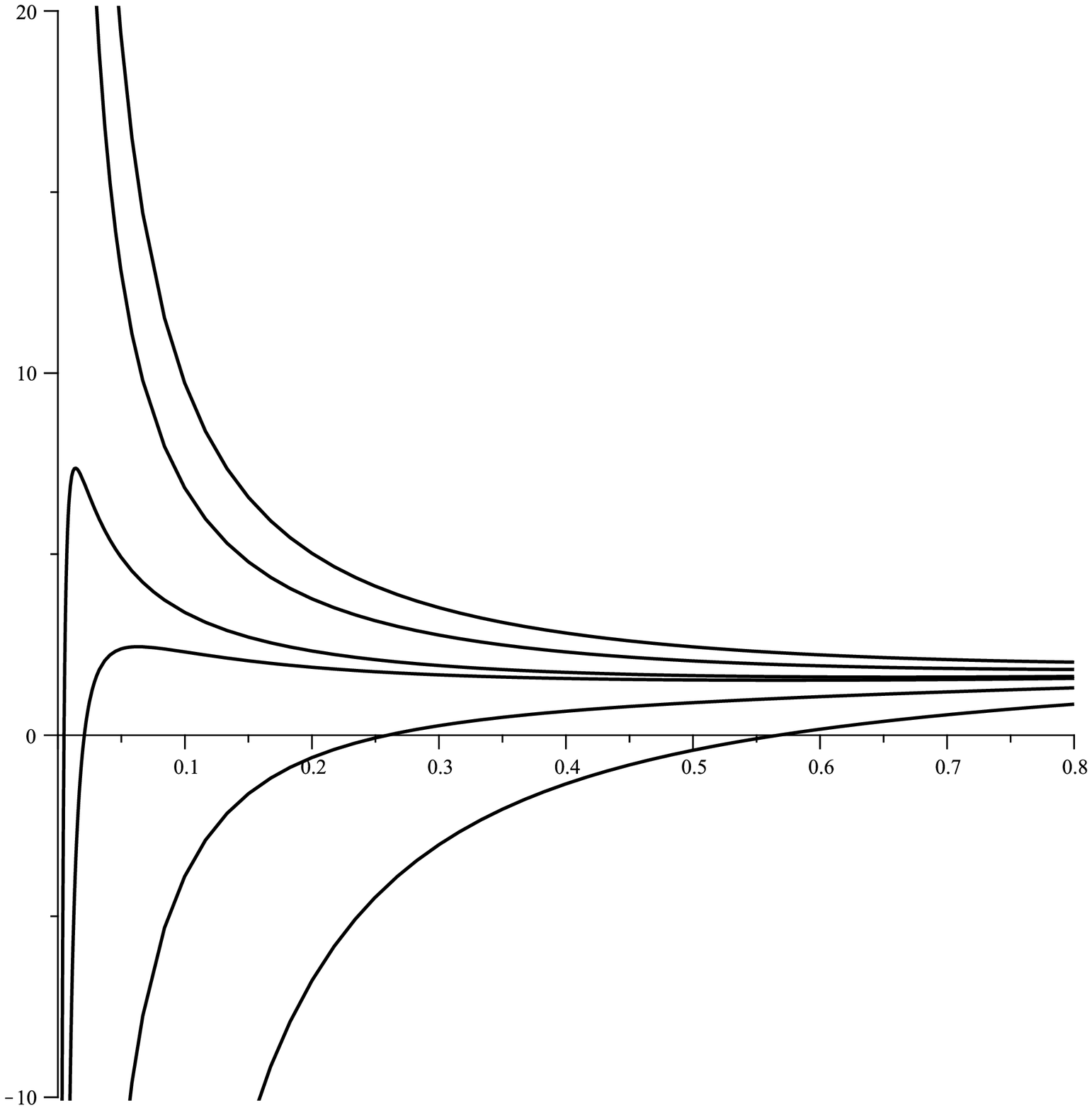} & \epsfxsize=8cm \epsffile{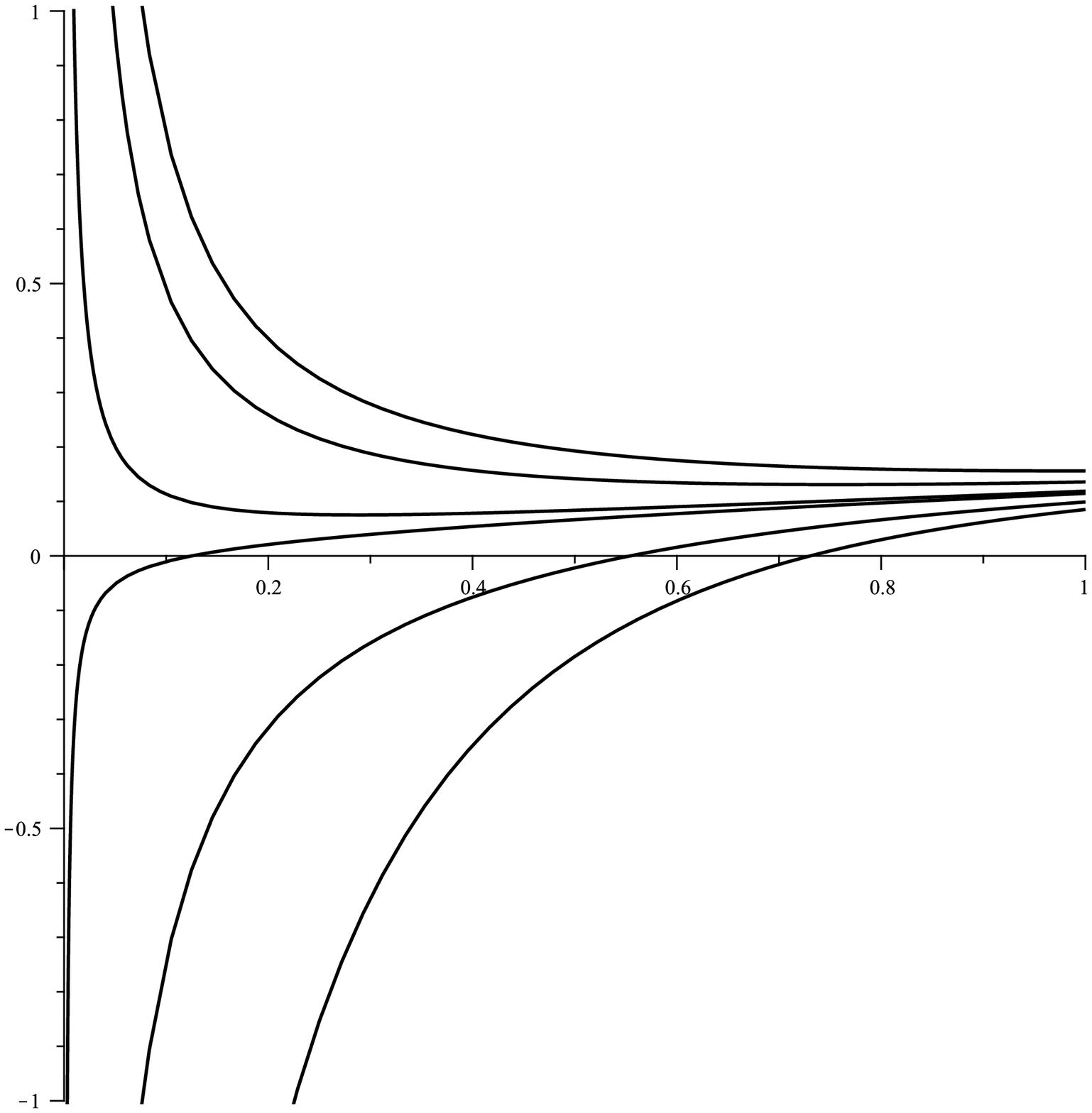}%
\end{array}
$%
\caption{Hawking temperature versus $r_{+}$ for $k=1$, $Q=1$, $\Lambda=-1$
and $\protect\beta=0.01,0.1,0.22,0.26,0.5,1.2$ from top to down. \emph{ENEF
(left figure) and LNEF (right figure)}}
\label{FigT}
\end{figure}

In order to determine the Hawking temperature of the black hole, we use the
enforcing regularity of the Euclidean section of the spacetime at the event
horizon, $r=r_{+}$, yielding
\begin{equation}
T=\frac{k-\Lambda r_{+}^{2}}{4\pi r_{+}}+\left\{
\begin{array}{ll}
\frac{\beta Q\left( 1-L_{W_{+}}\right) }{4\pi r_{+}\sqrt{L_{W_{+}}}}-\frac{%
\beta ^{2}r_{+}}{8\pi }, & \text{ ENEF}\vspace{0.2cm} \\
\frac{Q^{2}\left( 2-\Gamma _{+}\right) }{\pi r_{+}^{3}\Gamma _{+}\left(
\Gamma _{+}-1\right) }+\frac{\beta ^{2}r_{+}}{\pi }\left( \ln (\frac{\Gamma
_{+}^{2}-1}{2})-\frac{2}{\Gamma _{+}}\right) , & \text{LNEF}%
\end{array}%
\right. ,  \label{Temp}
\end{equation}%
where $\Gamma _{+}=\sqrt{1+\frac{Q^{2}}{r_{+}^{4}\beta ^{2}}}$ and $%
L_{W_{+}}=LambertW\left( \frac{4Q^{2}}{\beta ^{2}r_{+}^{4}}\right)$. In
order to find the effects of nonlinearity on the temperature, we plot $T$
for different values of $\beta$. Fig. \ref{FigT} shows that there is a
critical nonlinearity parameter, $\beta_{c}$, in which the Hawking
temperature is positive definite for $\beta<\beta_{c}$. This situation
appears for the black holes with one non-extreme horizon as it happens for
Schwarzschild solutions (for more details see \cite{HendiAnn}). Moreover, we
find that for $\beta>\beta_{c}$, there is a lower limit for the horizon
radius, $r_{0}$, in which $T$ will be positive for $r_{+} > r_{0}$.

Now, we are in a position to check the first law of thermodynamics. To do
this, we obtain the total mass, $M$, as a function of the extensive
quantities $S$ and $Q$
\begin{equation}
M(S,Q)=\frac{kS^{1/2}}{2\pi ^{1/2}}-\frac{\Lambda S^{3/2}}{6\pi ^{3/2}}%
+\left\{
\begin{array}{ll}
\frac{\beta Q}{6}\sqrt{\frac{S}{\pi \Pi }}\left[ 1+\Pi +\frac{4\Pi ^{2}}{5}%
\digamma {\left( \left[ 1\right] ,\left[ \frac{9}{4}\right] ,\frac{\Pi }{4}%
\right) }\right] , & \text{ENEF}\vspace{0.2cm} \\
\frac{8Q^{2}\digamma {\left( \left[ \frac{1}{2},\frac{1}{4}\right] ,\left[
\frac{5}{4}\right] ,1-\Upsilon ^{2}\right) }}{9\left( \frac{S}{\pi }\right)
^{1/2}}-\frac{2\beta ^{2}S^{3/2}\left( \ln \left[ \frac{2}{1+\Upsilon }%
\right] -\frac{5}{3}\left( 1-\Upsilon \right) \right) }{3\pi ^{3/2}}, &
\text{LNEF}%
\end{array}%
\right. ,  \label{Msmar}
\end{equation}
where $\Upsilon =\sqrt{1+\left( \frac{\pi Q}{S\beta }\right) ^{2}}$\ and $%
\Pi =LambertW\left( \frac{4\pi ^{2}Q^{2}}{\beta ^{2}S^{2}}\right) $.

Now, we can define the intensive parameters $T$ and $U$, conjugate to
extensive quantities $S$ and $Q$, respectively. These intensive quantities
are
\begin{equation}
T=\left( \frac{\partial M}{\partial S}\right) _{Q},\ \ U=\left( \frac{%
\partial M}{\partial Q}\right) _{S},  \label{Dsmar}
\end{equation}%
After some manipulation (with numerical calculations), one can show that the
temperature and electric potential calculated by Eq. (\ref{Dsmar}) coincide
with Eqs. (\ref{Temp}) and (\ref{U}), exactly. Eq. (\ref{Dsmar}) is nothing
but the first law of thermodynamics and therefore, these intensive and
extensive quantities satisfy the first law
\begin{equation}
dM=TdS+UdQ.  \label{1stLaw}
\end{equation}

\subsection{Thermodynamic stability}

Here, we investigate the thermal stability of the solutions. Depending on
the set of thermodynamic variable or state functions of a system, we can
examine the the thermodynamic stability from different point of views.
Usually, in the canonical ensemble, in order to study the thermodynamic
stability of the black holes with respect to small variations of the
thermodynamic coordinates, it is common to analyze the sign of the heat
capacity at constant electric charge
\[
C_{Q}\equiv T\left( \frac{\partial S}{\partial T}\right) _{Q}=T\left( \frac{%
\partial ^{2}M}{\partial S^{2}}\right) _{Q}^{-1}.
\]

The positivity of the $C_{Q}$ is a sufficient condition for the system to be
locally stable \cite{Chamblin,Davies}. In other word, the local stability of
a black hole will be ensured only if there exists a range of the event
horizon radius for which the heat capacity is positive.

\begin{figure}[tbp]
$%
\begin{array}{cc}
\epsfxsize=8cm \epsffile{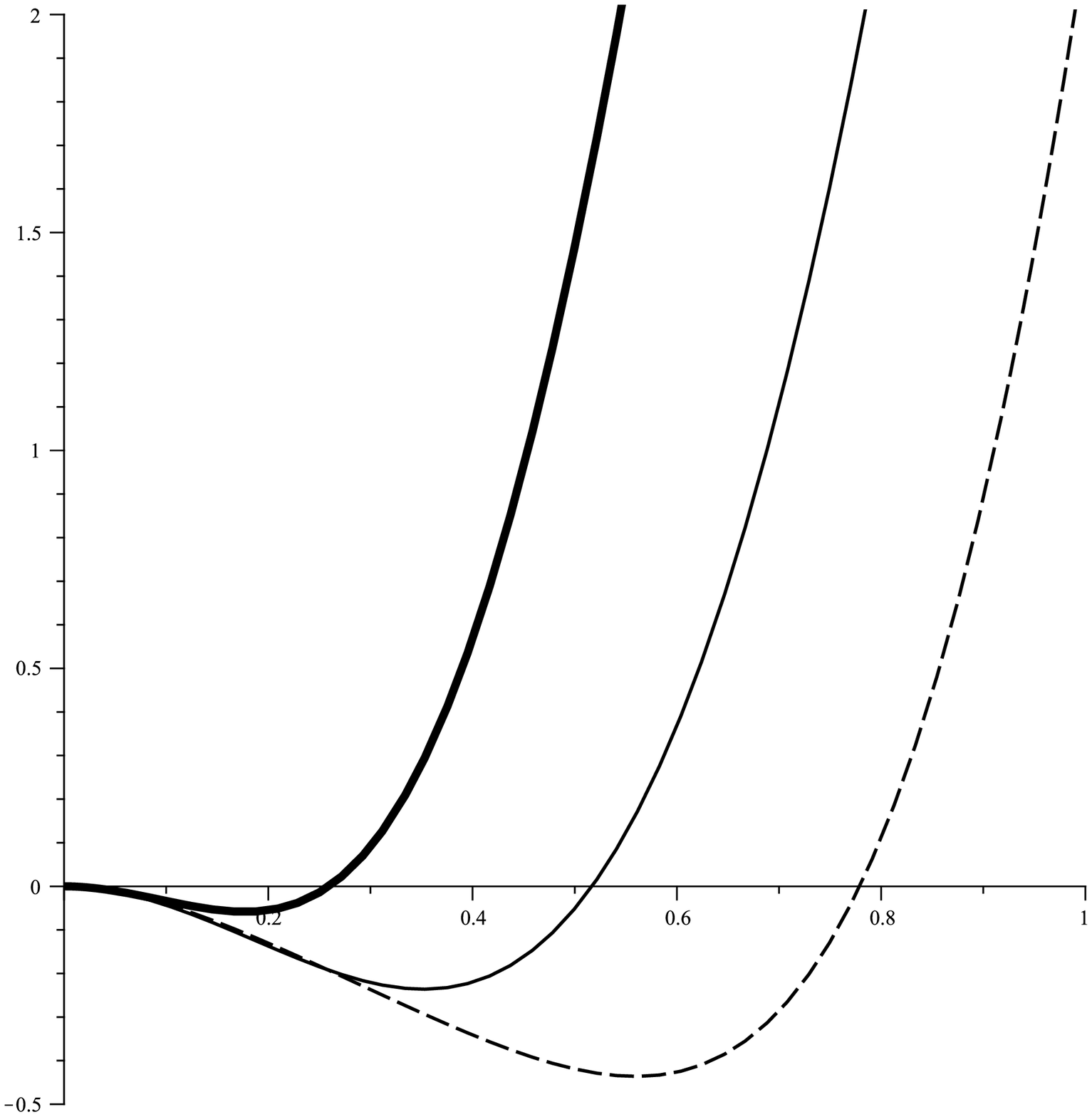} & \epsfxsize=8cm \epsffile{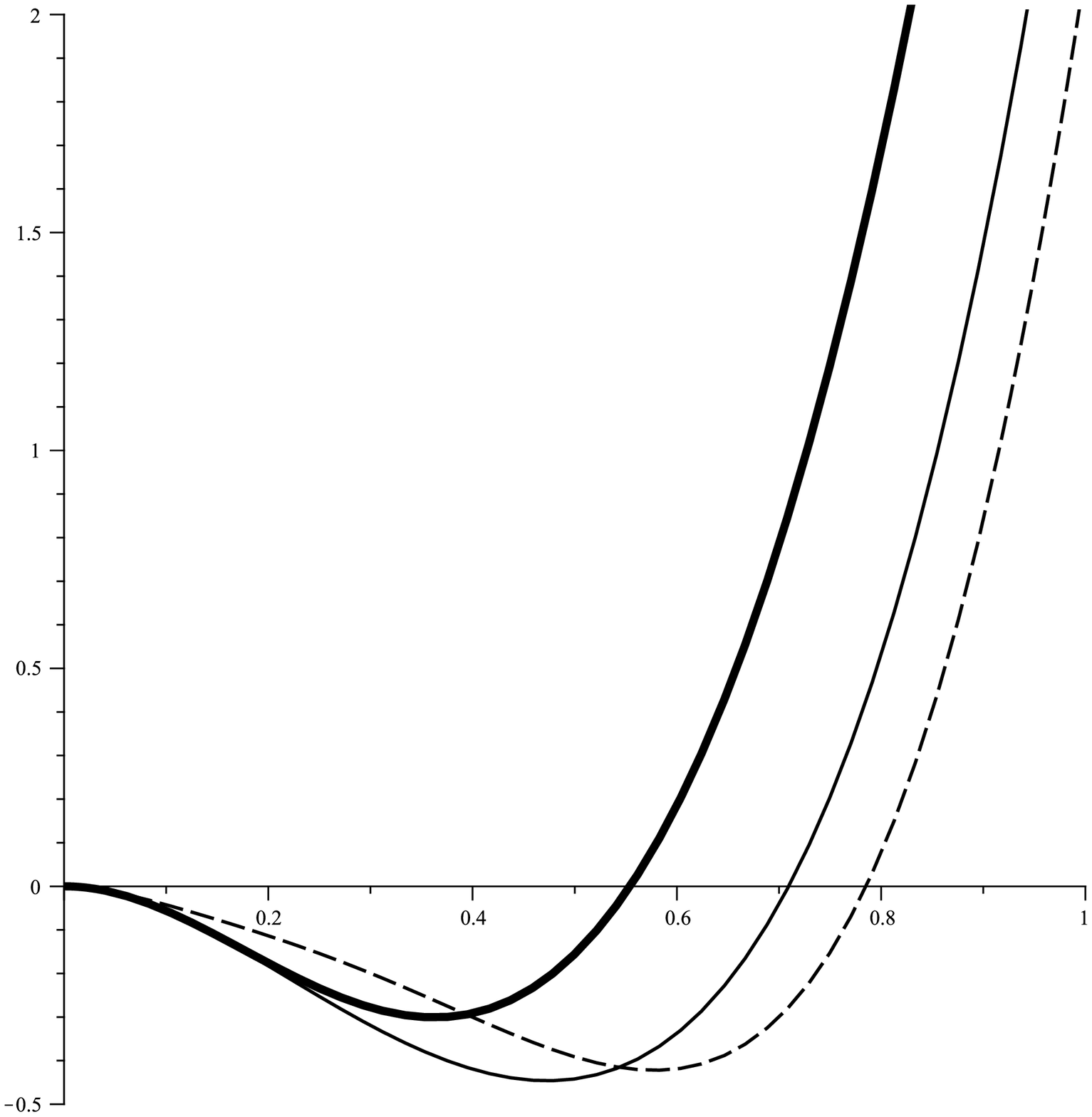}%
\end{array}
$%
\caption{Heat capacity versus $r_{+}$ for $k=1$, $Q=1$, $\Lambda =-1$ and $%
\protect\beta =0.5$ (bold line), $\protect\beta =1$ (solid line) and $%
\protect\beta =10$ (dashed line). \emph{ENEF (left figure) and LNEF (right
figure)}}
\label{HEAT}
\end{figure}

Using Eqs. (\ref{Entropy}), (\ref{Temp}) and (\ref{Msmar}), and after some
delicate simplifications, we obtain
\begin{equation}
C_{Q}=\left\{
\begin{array}{ll}
2\pi r_{+}^{2}\left( \frac{\left[ \Lambda r_{+}^{2}-k+\frac{\beta
^{2}r_{+}^{2}}{2}\right] \sqrt{L_{W_{+}}}-Q\beta \left( 1-L_{W_{+}}\right) }{%
\left[ \Lambda r_{+}^{2}+k+\frac{\beta ^{2}r_{+}^{2}}{2}\right] \sqrt{%
L_{W_{+}}}-Q\beta \left( 1+L_{W_{+}}\right) }\right) , & \text{ENEF}\vspace{%
0.2cm} \\
2\pi r_{+}^{2}\left( \frac{4\beta ^{2}\left[ \ln \left( \frac{1+\Gamma _{+}}{%
2}\right) +1-\Gamma _{+}\right] -\Lambda +kr_{+}^{-2}}{4\beta ^{2}\left[ \ln
\left( \frac{1+\Gamma _{+}}{2}\right) -1+\Gamma _{+}\right] -\Lambda
-kr_{+}^{-2}}\right) , & \text{LNEF}%
\end{array}%
\right. .  \label{CQ}
\end{equation}%
Numerical calculations show that there is a lower limit horizon radius for
stable solutions (see Fig. \ref{HEAT}). In other words, there is an $r_{min}$
in which for $r_{+}>r_{min}$ the heat capacity is positive and so the
solutions are stable. It is easy to find that the value of $r_{min}$ depends
on the values of the metric parameters and specially the nonlinearity
parameter. Fig. \ref{HEAT} shows that for fixed $Q$, $k$ and $\Lambda$, when
the value of the nonlinearity increases, the lower limit horizon radius
increases, too, which means that the nonlinearity parameter affects the
minimum size of stable black holes. Furthermore, in order to find the
effects of NLED, one can use the series expansion of the heat capacity for
large values of $\beta$, yields
\begin{equation}
C_{Q}=C_{EM}-\frac{\chi \pi Q^{4}\left( 4\Lambda
r_{+}^{4}+2Q^{2}-3kr_{+}^{2}\right) }{2r_{+}^{2}\left( kr_{+}^{2}+\Lambda
r_{+}^{4}-3Q^{2}\right) ^{2}}\beta ^{-2}+O(\beta ^{-4}),  \label{CQ2}
\end{equation}%
where $C_{EM}$ is the heat capacity of Einstein Maxwell gravity with the
following explicit form
\[
C_{EM}=\frac{2\pi r_{+}^{2}\left( \Lambda r_{+}^{4}+Q^{2}-kr_{+}^{2}\right)
}{4\left( \Lambda r_{+}^{4}-3Q^{2}+kr_{+}^{2}\right) },
\]%
and $\chi $ is equal to $8$ and $1$ for ENEF and LNEF branches,
respectively. In Eq. (\ref{CQ2}), one finds that the second term is the
leading NLED correction to the Einstein--Maxwell black hole solutions.

\subsection{Geometrothermodynamics}

In order to describe phase transitions of thermodynamic systems,
one can use the concept of geometry in thermodynamics (GTD) and
investigate curvature singularities so that the curvature can be
interpreted as a system interaction. This method was first
introduced by Weinhold \cite{Weinhold1,Weinhold2} whose Riemannian
metric defined as the second derivatives of internal energy with
respect to entropy and other extensive quantities (electric
charge) of a thermodynamic system. After that, Ruppeiner
\cite{Ruppeiner1,Ruppeiner2} introduced another metric, defined as
the negative Hessian of entropy with respect to the internal
energy and other extensive quantities of a thermodynamic system.
We should note that the Ruppeiner metric is conformally related to
the Weinhold metric with the inverse temperature as the conformal
factor \cite{ERconformal}.

\begin{figure}[tbp]
$%
\begin{array}{cc}
\epsfxsize=8cm \epsffile{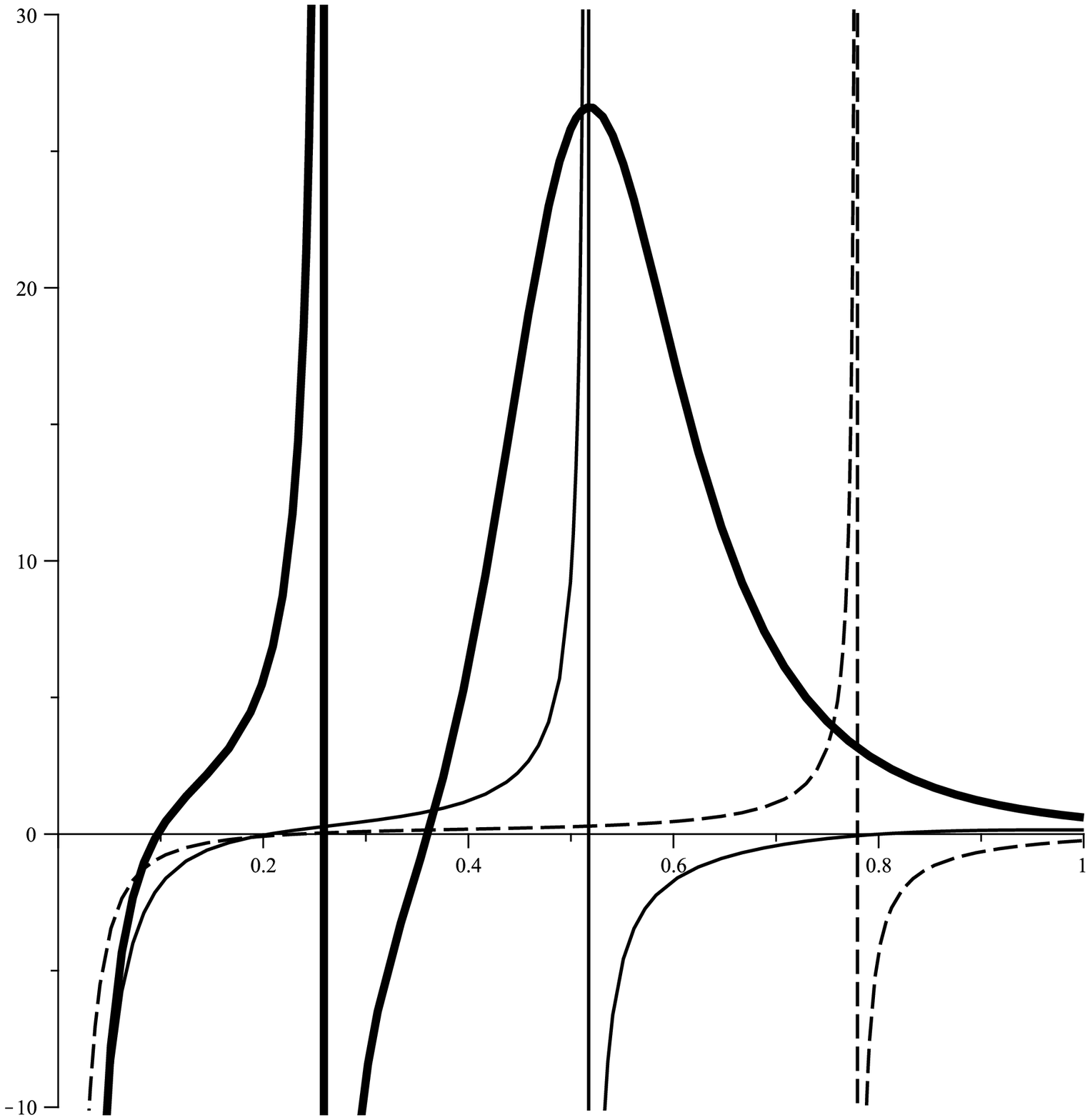} & \epsfxsize=8cm \epsffile{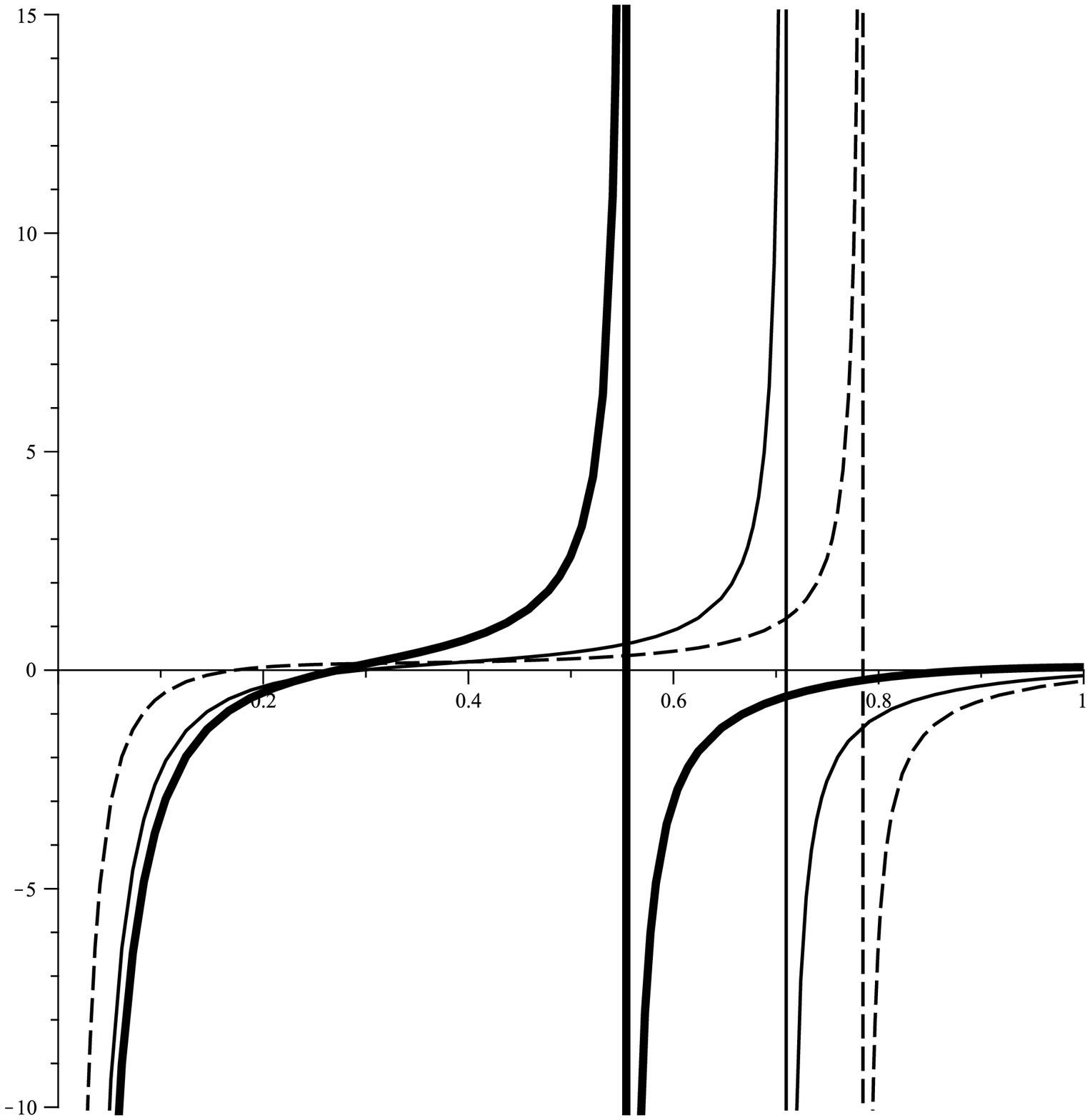}%
\end{array}
$%
\caption{Ricci scalar of GTD method versus $r_{+}$ for $k=1$,
$Q=1$, $\Lambda =-1$ and $\protect\beta =0.5$ (bold line),
$\protect\beta =1$ (solid line) and $\protect\beta =10$ (dashed
line). \emph{ENEF (left figure) and LNEF (right figure)}}
\label{GTDfig}
\end{figure}

Sometimes, the singular points of the Weinhold and Ruppeiner
metrics are not consistent with the ones of the heat capacity,
unfortunately. In order to solve this puzzle, one can use the
Quevedo method
\cite{Quevedo1,Quevedo2,Quevedo3,Quevedo4,Quevedo5,Quevedo6,Quevedo7},
whose proposed metric is Legendre-invariant. As we know, there are
many Legendre-invariant metrics that one can use. Here, we can use
the simplest Legendre invariant generalization of Weinhold's
metric, $g^{W}$, can be written in components as
\begin{equation}
g=Mg^{W}=M\frac{\partial ^{2}M}{\partial X^{a}\partial X^{b}}dX^{a}dX^{b},
\label{Wein}
\end{equation}%
where $X^{a}=\left\{ S,Q\right\} $ and its Legendre invariant metric can be
written in terms of the Ruppeiner's metric, $g^{R}$, as%
\begin{equation}
g=MTg^{R}=-\frac{M}{\left( \frac{\partial S}{\partial M}\right) }\frac{%
\partial ^{2}S}{\partial Y^{a}\partial Y^{b}}dY^{a}dY^{b},  \label{Rupp}
\end{equation}%
where $Y^{a}=\left\{ M,Q\right\} $. Considering Eq. (\ref{Msmar}),
we use the Legendre invariant of Ruppeiner's metric to calculate
the curvature scalar. Although the method is straightforward, for
reason of economy, we do not present the analytical long equations
of the Ricci scalars. We plot $R(S,Q)$ as a function of $r_{+}$
($r_{+}=\sqrt{S/\pi }$) to compare with Fig. \ref{HEAT}. Comparing
Figs. \ref{GTDfig} and \ref{HEAT}, we find that the singularities
of the Ricci scalar ( Fig. \ref{GTDfig}) take place at those
points where the heat capacity vanishes (Fig. \ref{HEAT}) and the
black hole undergoes a phase transition.

\section{Generalization to higher dimensions \label{d-dim}}

Here, we generalize our solutions to $d$-dimensional ones. Considering the
following metric
\begin{equation}
ds^{2}=-f(r)dt^{2}+\frac{dr^{2}}{f(r)}+r^{2}d\hat{g}_{k}^{2},  \label{met2}
\end{equation}%
where $d\hat{g}_{k}^{2}$ is the metric of ($d-2$)-dimensional hypersurface
of constant curvature $(d-2)(d-3)k$ with the following explicit form%
\begin{equation}
d\hat{g}_{k}^{2}=\left\{
\begin{array}{cc}
d\theta _{1}^{2}+\sum\limits_{i=2}^{d-2}\prod\limits_{j=1}^{i-1}\sin
^{2}\theta _{j}d\theta _{i}^{2} & k=1 \\
d\theta _{1}^{2}+\sinh ^{2}\theta _{1}d\theta _{2}^{2}+\sinh ^{2}\theta
_{1}\sum\limits_{i=3}^{d-2}\prod\limits_{j=2}^{i-1}\sin ^{2}\theta
_{j}d\theta _{i}^{2} & k=-1 \\
\sum\limits_{i=1}^{d-2}d\phi _{i}^{2} & k=0%
\end{array}%
\right. .  \label{dOmega2}
\end{equation}

Taking into account $d$-dimensional electromagnetic and gravitational field
equations, we find that the nonzero components of the electromagnetic field
are
\begin{equation}
F_{tr}=-F_{rt}=\frac{Q}{r^{d-2}}\times \left\{
\begin{array}{cc}
\exp \left( -\frac{L_{W_{d}}}{2}\right) , & ENEF \\
\frac{2}{\Gamma _{d}+1}, & LNEF%
\end{array}%
\right. ,  \label{h(r)D}
\end{equation}%
where
\begin{eqnarray*}
L_{W_{d}} &=&LambertW\left( \frac{4Q^{2}}{\beta ^{2}r^{2d-4}}\right) , \\
\Gamma _{d} &=&\sqrt{1+\frac{Q^{2}}{\beta ^{2}r^{2d-4}}},
\end{eqnarray*}
and $d$-dimensional metric function may be written as
\begin{equation}
f(r)=k-\frac{2\Lambda r^{2}}{(d-1)(d-2)}-\frac{M}{r^{d-3}}+\Sigma ,
\label{f(r)D}
\end{equation}%
where%
\begin{equation}
\Sigma =\left\{
\begin{array}{cc}
-\frac{\beta ^{2}r^{2}}{(d-1)(d-2)}-\frac{2Q\beta }{(d-2)r^{d-3}}\int \frac{%
L_{W_{d}}-1}{\sqrt{L_{W_{d}}}}dr, & ENEF \\
-\frac{16\beta ^{2}r^{2}}{(d-1)(d-2)}-\frac{8\beta ^{2}\ln (2)}{(d-1)(d-2)}+%
\frac{8}{(d-2)r^{d-3}}\int \frac{\frac{Q^{2}}{\Gamma _{d}-1}-\beta ^{2}\ln
\left( \frac{\beta ^{2}r^{2d-4}\left( \Gamma _{d}-1\right) }{Q^{2}}\right) }{%
r^{d-2}}dr, & LNEF%
\end{array}%
\right. .  \label{sigma}
\end{equation}

\section{Conclusions}

In this paper, we have considered Einstein gravity coupled to NLED in the
form of exponential and logarithmic in the Lagrangian of the matter field.

Obtained solutions are generalization of topological black holes in the
Einstein--Maxwell gravity and, as one expected, these solutions reduce to
topological Reissner--Nordstr\"{o}m black hole for large values of the
nonlinearity parameter, $\beta$. Regarding a finite and fix value for $\beta$%
, one can find that the asymptotic behavior of the obtained solutions are
the same as those in asymptotically adS topological Reissner--Nordstr\"{o}m
black holes.

The main goal of this paper is analyzing the thermodynamics properties of
the black hole solutions. Hence, we have calculated the conserved and the
thermodynamic quantities and check the validity of the first law of
thermodynamics. Then, we studied the thermodynamic stability of the
solutions and found that there is a lower limit for the size of stable black
holes. We have plotted the heat capacity with respect to $r_{+}$ and showed
that when $\beta $ increases, the minimum size of stable black holes
increases, too.

We also applied GTD approach proposed by Quevedo whose metric is
Legendre-invariant. We used the Legendre invariant of Ruppeiner's
metric to calculate its curvature scalar and found that the
singularities of the Ricci scalar in GTD method take place at
those points where the heat capacity vanishes, which is the black
hole phase transition.

Last section is devoted to generalization of four dimension black holes to
higher dimensional solutions. It will be straightforward to investigate
thermodynamic properties of higher dimensional black holes and for reason of
economy, we did not give their details.

Finally, it is worthwhile to mention that it would be interesting
to generalize these Einsteinian static solutions to higher
derivative gravity and rotating solution. In addition, one can
investigate the non-strictly thermality of the Hawking radiation
spectrum \cite{Parikh}, quasi-normal modes and the Corda effective
state of black holes \cite{Corda1,Corda2,Corda3,Corda4,Corda5}
with considering the emission of Hawking quanta as a discrete
process rather than a continuous process. We leave these problems
for future works.

\begin{acknowledgements}
We thank Shiraz University Research Council. This work has been
supported financially by Research Institute for Astronomy \&
Astrophysics of Maragha (RIAAM), Iran.
\end{acknowledgements}


\end{document}